\documentclass[12pt]{article}
\newcommand{\be}{\begin{equation}}
\newcommand{\ee}{\end{equation}}
\newcommand{\ba}{\begin{eqnarray}}
\newcommand{\ea}{\end{eqnarray}}
\newcommand{\no}{\noindent}
\newcommand{\n}{\label}

\newcommand{\kf}{$k$-field }
\usepackage[notref,notcite]{showkeys}
\usepackage{graphicx}
\usepackage{bm}

\begin{document}

\title{Atypical k-essence cosmologies}
\author{ \large Luis P. Chimento$^1$\footnote{chimento@df.uba.ar} \addtocounter{footnote}{5}  and Ruth Lazkoz $^2$\footnote{wtplasar@lg.ehu.es}\\
{\it \small Dpto. de F\'\i sica, Facultad de Ciencias Exactas y Naturales, }\\{\it \small Universidad de Buenos Aires, Ciudad Universitaria,}\\
{\it \small  Pabell\'on I, 1428 Buenos Aires, Argentina}\\
{\it  $^1$ \small Fisika Teorikoa eta Zientziaren Historiaren Saila, }\\{\it \small Zientzia eta Teknologiaren Fakultatea, Euskal Herriko Unibertsitatea,}\\
{\it \small  644 Posta Kutxatila, 48080 Bilbao, Spain}}

\maketitle
\begin{abstract}

We analyze the implications of having a divergent speed of sound in k-essence
cosmological models. We first study a known theory  of that kind, for which
the Lagrangian density depends linearly on the time derivative of the k-field.
We show that when  k-essence is the only source consistency requires that the potential of the k-field be of the
inverse square form. Then, we review the  known result that the corresponding
power-law solutions can be mapped to power-law solutions of theories with no
divergence in the speed of sound. After that, we argue that the requirement of
a divergent sound speed at some point fixes uniquely the form of the
Lagrangian  to be exactly the one considered earlier and prove the asymptotic
stability of  the most interesting solutions belonging to the divergent
theory. Then, we discuss the implications having not just k-essence but also matter. This is
interesting because  introducing another component breaks the
rigidity of the theory, and the form of the potential ceases to be unique as happened in the pure k-essence case.
Finally, we show the finiteness of the effective sound speed under an appropiate definition.
\end{abstract}

\newpage
\section{Introduction}

Mainstream models of accelerated expansion in the universe assume it is due to
the dynamics of scalar fields evolving in a self-interaction potential. In
general, the Lagrangians of the effective theories describing those fields
include  non-canonical kinetic terms, which might be responsible for crucial
cosmological consequences like the occurrence of inflation even without a
potential (purely kinetic acceleration or k-acceleration)
\cite{kinflation,scherrer,Chimento}. In these models, inflation is pole-like, that is, the scale factor evolves like a negative power of time.
An earlier theoretical framework in which
(pole-like) k-acceleration arises naturally is the pre-big bang model of  string cosmology \cite{prebigbang}. In this setup, acceleration  is just due to a scalar field called the dilaton, and it will only manifest itself in the string conformal frame. Finally, for other ideas on kinetic inflation one may have a look at \cite{levin}, where acceleration was put down to a dynamical Planck mass.

Coming back to k-essence, the non-canonical terms considered in the Lagrangian will only be combinations of the square of the gradient of the scalar field (hereafter k-field), because the equations of motion in  classical theories seem to be of second order. Moreover,
since k-fields can be used for constructing dark energy models it is common
place to interpret them as some kind of matter called k-essence
\cite{scherrer,k-essence,tracking}. Nevertheless,  originally k-fields were
not introduced for the  description of late time acceleration, but rather they
were suggested  as possible  inflation driving agents
\cite{kinflation,kinfper}. Lately, efforts in the framework of k-essence have
been directed toward model building using  power-law solutions which preserve
\cite{Chimento,tacpow,ChiFein} or violate the weak-energy condition
\cite{AguChiLaz}.

In this paper, we revisit k-essence cosmologies with an infinite sound speed,
and throw in more light on their implications. In Section 2 we discuss the
main features of the models and prove that for consistency  the potential must
be of inverse square form. In Section 3 we construct models with a divergent
speed of sound and we show that the form of the Lagrangian giving rise to an
infinite speed of sound is uniquely determined. We also provide an alternative
view on the origin of such models which relies on how the Hubble factor
depends on the k-field and its derivative. At the end of this section the
asymptotic stability of the most appealing solutions within that framework is
investigated. In Section 4 we consider a more general model to include matter
together with the k-essence, we study the implications of this generalization, 
and then we perturb the background geometry and find the
effective sound speed. Then, for illustration we calculated for a simple cosmological model
in which the batrotropic index has a logarithmic dependence on the k-field.
Finally, in Section 5 we draw our main conclusions.

\section{Main features of the models}

In usual practice, k-essence is defined as a scalar field $\phi$ with non-canonical kinetic energy associated with a factorizable Lagrangian of the form
\begin{equation}
{\cal L}\equiv V(\phi)F(x), 
\end{equation}where 
$
x={\nabla_{\mu}{\phi}\nabla^{\mu}{\phi}}
$
and $F(x)$ is a function of the kinetic energy $x$.
This form of the Lagrangian is suggested by the Born-Infeld  one
 \begin{equation}
{\cal L}=-V(\phi)\sqrt{1+x},
\end{equation} which was associated with the tachyon  by computations in boundary string
field theory \cite{BSFT}. Such Lagrangian also arises in open bosonic string theory \cite{fratse} and is a key ingredient in the effective
theory of D-branes \cite{Leigh}.

Using the perfect fluid analogy, the energy density and the pressure are given by
{\setlength{\arraycolsep}{1pt}
\begin{eqnarray}
\rho_\phi&=&V(F-2xF_x),\\
p_\phi&=&-VF.
\end{eqnarray}}
 We assume from now on a flat Friedmann-Robertson-Walker (FRW) spacetime
with line-element
\begin{equation}
\label{metric}
ds^{2} =-dt^{2}+
a^{2}(t)\left(dx^{2}_{1}+dx^{2}_{2}+dx^{2}_{3}\right),
\end{equation}
 where $a(t)$ is the scale factor and $\phi$ is homogeneous so that
$x=-\dot\phi^2$. Let $H\equiv\dot a/a$ be the Hubble factor, then  the Einstein equations reduce to
{\setlength{\arraycolsep}{1pt}
\begin{eqnarray}
&&3H^2=\rho_\phi, \n{00} \\
&&\dot H=xVF_x \n{11},
\end{eqnarray}
where we have taken units such that $8\pi G=1$.
A consequence of the latter is the conservation equation
\begin{equation}
\n{kg}
(F_x+2xF_{xx})\ddot\phi+3HF_x\dot\phi+\frac{V'}{2V}\,(F-2xF_x)=0.
\end{equation}
Furthermore, if we write the  equation of state in the form
$p_\phi=(\gamma_\phi-1)\rho_\phi$, the barotropic  index
$\gamma_\phi$ will read
\begin{equation}
\n{ga}
\gamma_\phi=-\frac{2\dot H}{3H^2}=-\frac{2xF_x}{F-2xF_x}
\end{equation}

In \cite{ChiFein} different classes of FRW k-essence cosmologies were
investigated. Among them, those which lead to power-law  solutions with an inverse
square potential and  scalar field evolving linearly with time were analyzed,
and they were shown to be related by a one-to-one map (see next section) to  power-law solutions
arising from a theory with
\be
F=\alpha+\beta\sqrt{-x}\label{divergent},
\ee

\no  where $\alpha$ and $\beta$ are arbitrary constants. Interestingly, the
latter is a particular case of the $F$ associated with the extended tachyon
models considered in \cite{Chimento}. There  solutions were found with the {\it a priori}
assumption that the potential should be of 
 inverse square type. In contrast, we will demonstrate below that, in fact, there is no other possibility.
 
Now, the cosmological models one obtains from (\ref{divergent}) can  be viewed as having an infinite
sound speed if  the familiar definition
\begin{equation}
\n{cs}
c_s^2=\frac{p_{\phi x}}{\rho_{\phi x}}=\frac{F_x}{F_x+2xF_{xx}}
\end{equation}  
is used. For that reason, we will refer here to  the theory arising from
($\ref{divergent}$) as the divergent theory, and, by opposition, any other
theory for which $c_s\ne\infty$ will be labeled as non-divergent.

As we will show immediately, another peculiarity of the theory is that
compatibility  requires the potential be of the inverse square form.  For our choice of
$F$ the $k$-field equation (\ref{kg}) becomes
 \begin{equation}
 3HF_x\dot\phi+\frac{V'}{2V}\,(F-2xF_x)=0\label{kfieldeq},
 \end{equation}
Now, using the definitions of   $x$ and $\gamma_\phi$ once and twice respectively
one arrives at
\begin{equation}
\n{V.}
 \frac{\dot V}{V}=\frac{2\dot H}{H},
\end{equation} 
which gives
$
V\propto H^2$.
On the other hand, inserting (\ref{divergent}) in (\ref{ga}), one gets

\be
\n{ga1}
\gamma_\phi=-\frac{\beta}{\alpha} \dot\phi,
\ee
which after integration leads to
\begin{equation}
H\phi=\frac{1}{\Gamma_0},\label{Hphi}
\end{equation} with $\Gamma_0=-3\beta/2\alpha$, so that  we can finally write
\begin{equation}
V=\frac{V_0}{\phi^2},\label{phisquare}
\end{equation}
 where $V_0=4\alpha/3\beta^2$ is a constant. Note also that we have set the origin of the potential
at $\phi=0$. Summarizing, one can view the result as  if the simultaneous
requirement that $H\phi\equiv\rm{constant}$ and that $V$ is of the inverse
square form characterized the solutions to the $k$-field equation
(\ref{kfieldeq}) for an $F$ like (\ref{divergent}).

At this stage, we can insert (\ref{divergent}) into the Einstein equations and use  
(\ref{Hphi}) and (\ref{phisquare}). The information we extract is that, necessarily,
\begin{equation}
\n{f}
F=\frac{3H^2\phi^2}{V_0}-\frac{2H\phi}{V_0}\sqrt{-x},
\end{equation}
where one should keep in mind that $H\phi$ must be replaced by $\Gamma_0^{-1}$.

Remarkably, since   there is no  evolution equation for $\phi$ in this theory,  the time dependence
of $\phi$ or $H$ are not fixed by the form of the potential, and solutions belonging to this theory
exist for absolutely any evolution one can imagine.
\section{Models with divergent speed of sound}

Consider now  theories with $F$ functions different from (\ref{divergent})
 and their power-law solutions, which are obtained under the hypotheses  
\begin{eqnarray}
\n{pl}
V=\frac{V_0}{\phi^2}, \\
  \phi=\phi_0\,{t},
\end{eqnarray} which clearly imply $F$ is constant. Although these particular models
arise from non-divergent theories, they share the property $H\phi=\rm{constant}$ with  all the  models derived from the divergent theory, and in particular with the power-law ones. Thus, convenient choices of the free parameters will allow for one-to-one maps between power-law solutions of the divergent and non-divergent theories.

Specifically, if for
the cosmologies arising from the non-divergent theories we set 
$a=a_0t^n$ with $a_0$ a constant, we will have $H\phi=n\phi_0$, and the isomorphism will follow from the requirement $n\phi_0=\Gamma_0^{-1}$, which trough the k-field equation 
(\ref{kfieldeq}) enforces
\begin{eqnarray}
\alpha=\frac{3n^2\phi^2_0}{V_0}, \label{alpha}\\
\beta=-\frac{2n\phi_0}{V_0}. \label{beta}
\end{eqnarray}

Nevertheless, despite this equivalence argumentation, the models arising
from the divergent and non-divergent theories are not completely 
interchangeable in all respects. Before we go deeper into this matter, it is
convenient to  introduce the parameters $f$ and $f'$, which respectively
stand for the function $F$ and its first derivative evaluated at
$x=x_0=-\phi_0^2$, that is,
\begin{eqnarray}
f=F(-\phi^2_0),\\
f'=F_x(-\phi^2_0).
\end{eqnarray} 
 If we substitute the latter into 
the Friedmann and $k$-field equations (\ref{00}) and (\ref{kg}) we find that the
index $n$ and the slope of the potential $V_0$ are given by
\begin{eqnarray}
\n{sol}
&&n=\frac{f+2\phi^2_0f'}{3\phi^2_0f'}\\
&&V_0=\frac{n}{f'},
\end{eqnarray} 
 
Now, inhomogeneous perturbations to the background FRW geometry would involve
the speed of sound $c_s$, or equivalently the second order derivative of $F$,
through equation (\ref{cs}). Thus, a measure of those perturbations would
provide information on $c_s$, which could be break the degeneracy of the
divergent theory and be used to restrict the set of admissible $F$ functions.
This provides an adequate framework where the effective sound speed can be
introduced, as it will be seen in Section 4.

Interestingly, there is a consistency argument that supports the validity of
the above result. In \cite{Chimento}  the first integral of
the $k$-field equation (\ref{kg}) for any $F$ expression was found provided the
coefficient of $\ddot \phi$ does not vanish; it reads

\begin{equation}
\n{fi}
\frac{\gamma_\phi}{\dot\phi}=\frac{1}{\phi}\left(\frac{2}{3H}+\frac{c}{a^3H^2}\right),
\end{equation}
with $c$ an arbitrary integration constant.
Consistency, nevertheless,  would require that (\ref{fi}) admitted as a particular result $H\phi=\rm{constant}$ which must otherwise hold in the limit in which the coefficient
of $\ddot \phi$ in the $k$-field equation. Recalling that in such case one must have
$\gamma_\phi=2\Gamma_0\dot\phi/3$,
the condition $H\phi=\Gamma_0^{-1}$ follows for $c=0$.

\subsection{Obtaining the divergent theory}

Let us now try to deepen the understanding of the implications of an infinite
sound speed. It is clear that an $F$ like (\ref{divergent}) (or  (\ref{f})) is associated with a divergent sound speed, but the question that comes to mind is whether such divergence could occur
for a different form of $F$. In order to find the answer, we are going to consider that the function $F$ is not
{\it a priori} of the form (\ref{divergent}), but rather just assume that the speed
of sound diverges at the point $x=x_0$, which means $(F'+2xF'')_{x=x_0}=0$ and
$\dot\phi=\dot\phi_0$. Using the $k$-field equation for the inverse square
potential $V=V_0/\phi^2$ recursively, we calculate below the values of $F$
and its derivatives at that point, i.e., $F(-x_0)=F(-\dot\phi_0^2)=F_0,
F'(-x_0)=F'_0$, and so on. Hence, at $x=x_0$, Eqs. (\ref{00}) and (\ref{kg})
become
\begin{eqnarray}
3H^2=V(F_0-2x_0F'_0), \n{000}\\
3HF'_0\dot\phi_0-\frac{F_0-2x_0F'_0}{\phi}=0.
\end{eqnarray}

\no Combining these equations with Eq. (\ref{11}), we arrive at

\begin{equation}
H\phi_0+\dot H\phi=0,
\end{equation}

\no and

\be
\n{f'}
F'_0=\frac{H\phi}{V_0\sqrt{-x_0}},
\ee

\no with $H\phi=\rm{constant}$, as can be seen from Eq. (\ref{000}). Besides, from the
vanishing of $(F'+2xF'')_{x=x_0}=0$, we get

\be
\n{f''}
F''_0= \frac{H\phi}{2V_0(-x_0)^{3/2}}.
\ee

\no Differentiating the \kf equation (\ref{kg}) recursively and using the above
results we obtain the remaining derivatives of the function $F$,

\be
\n{f'''}
F'''_0=\frac{3H\phi}{4V_0(-x_0)^{5/2}}
\ee

\no and so on. Precisely, these values of $F_0$, $F'_0$ ,..., coincide with those
obtained from the function (\ref{f}) and its derivatives evaluated at $x=x_0$. This shows
that the form of $F$ associated with an infinite speed of sound is unique, and it is necessarily given by (\ref{divergent}). In consequence, if the speed of sound is infinite
for some value of $x_0$, it will be so for every other value.

Although the result we just gave is quite strong,
we wish to present yet one more argument to shed
some more light on the origin of the divergent theory.
In principle, the Hubble factor $H$ might depend on the field $\phi$ and its first derivative $\dot\phi$ (no dependence on higher order derivatives is required because the k-field equation makes them all depend in turn on $\phi$ and $\dot\phi$). Now, let us assume for the time being that there is only
dependence on $\phi$, i.e.
\begin{equation}
H=h(\phi). 
\end{equation}
We can then calculate the barotropic index
\begin{equation}
\gamma_\phi=-\frac{2h'\dot\phi}{3h^2},
\end{equation}
which by definition (see Eq. (\ref{ga})) cannot depend on $\phi$, so that
\begin{equation}h'\propto h^2,
\end{equation} is required. From the latter it may be concluded that
 $h\propto\phi^{-1}$. Therefore,
$\gamma_\phi\propto\dot\phi$ and $H^2\propto\phi^{-2}$. This just means that
\begin{equation}
F-2xF_x\equiv\rm{constant},
\end{equation}
which remarkably solves to give the function $
F=\alpha+\beta\sqrt{-x}$
with arbitrary $\alpha$ and $\beta$.

Summarizing, just by making the hypothesis that $H$ depends on $\phi$ only, it is possible to
obtain the divergent theory without having to use the conservation equation. We feel
this clarifies why the divergent theory is atypical; it looks as if all k-essence theories
would split into two classes: the divergent class which is determined
by the latter  $F$, and the class of the theories generated by 
all the remaining  $F$ functions.

\subsection{Stability of the solutions}

Now, in order to go a fit further, we are going to study the stability
of the solution belonging to the divergent theory
against changes in the initial conditions.
Since it corresponds to $\gamma_\phi/\dot\phi=\rm{constant}$ we introduce a new variable \begin{equation}
\Gamma\equiv\frac{3\gamma_\phi}{2\dot\phi},
\end{equation}
so that  the unperturbed solution is represented by $\Gamma=\Gamma_0$, with
$\Gamma_0$ a constant. Let us write
now the $k$-field equation in terms of $\Gamma$ \cite{Chimento}:
\begin{equation}
\dot\Gamma+3\left(H \Gamma+\frac{V'}{2V}\right)(1-\gamma_\phi)=0.\label{kg_gamma}
\end{equation}
A constant solution $\Gamma=\Gamma_0$ to the last equation exists when
the condition
\begin{equation}
\n{cond}
H \Gamma_0+\frac{V'}{2V}=0, 
\end{equation}
holds, so that (\ref{kg_gamma}) can be cast as
\begin{equation}
\dot\Gamma+3H \Gamma\left(\Gamma-\Gamma_0\right)(1-\gamma_\phi)=0.\label{kg_gamma_bis}
\end{equation}
In addition, integrating the condition (\ref{cond}), we get the inverse square
potential (\ref{phisquare}) and the general solution of Eq. (\ref{kg_gamma})
is given by
\begin{equation}
\Gamma=\Gamma_0+\frac{m}{a^3H^2},
\end{equation}
where $m$ is an arbitrary integration constant. Therefore, the $\gamma_\phi>1$
solutions (and, in particular, all the accelerated ones) evolve asymptotically toward $\Gamma=\Gamma_0$.

A more subtle distinction between the models, which would   remove the
degeneracy, could be established by perturbing these solutions.  This,
however, is beyond the scope of the present paper.

\newpage

\section{Matter contribution}

We consider now a model with matter interacting with k-essence generated by
the kinetic function (\ref{divergent}). In addition, we assume the interaction
happens  through the geometry only, that is, the components are conserved
separately. In this case the Einstein equations are given by

\be
\n{00m}
3H^2=\alpha V+\rho_m,
\ee

\be
\n{ck}
-3\beta H+\alpha \frac{V'}{V}=0,
\ee

\be
\n{cm}
\dot\rho_m+3H(\rho_m+p_m)=0,
\ee

\no where Eqs. (\ref{ck}) y (\ref{cm}) express the conservation of k-essence and
matter respectively (Eqs. (\ref{00m}) and (\ref{ck}) follow from assuming Eq. (\ref{divergent})). Furthermore, if we write the  equation of state of the
matter in the form  $p_m=(\gamma_m-1)\rho_m$, then integrating the matter
conservation equation (\ref{cm}), we get $\rho_m=\rho_0/a^{3\gamma_m}$. On the
other hand, unlike in the case without matter, Eq. (\ref{ck}) cannot be
integrated, so we cannot say that the proportionality rule  $V\propto H^2$
found in Sec. II (ver Eq, (\ref{V.})) does not apply now. This is a
consequence of the fact that $\gamma_\phi=-2\dot H/3H^2$ is no longer valid,
that relation is rather satisfied by the overall barotropic index defined by

\be
\n{ov}
\gamma=\frac{\gamma_\phi\rho_\phi+\gamma_m\rho_m}{\rho_\phi+\rho_m}=-\frac{2\dot
H}{3H^2}.
\ee

This shows the potential is no longer fixed by the theory, and this represents
a crucial difference between the situation in which the universe is filled
with k-essence only, or with such fluid together with matter of some other
kind. Somehow matter modifies the results found in  Sec. II, thus allowing for
a more  realistic model, because the theory is not rigid anymore as when
only the k-essence is present, thus matter and k-essence with an arbitrary
potential jointly rule the cosmic dynamics.  Constructing a model matching the observations
and close enough to a Cosmological Constant should be quite easy by getting the potential sufficiently flat, and beta sufficiently small in
Eq.(43).

\subsection{The effective sound speed}

A realistic model should explain recent observations, which suggest that most
of the energy density of the universe consists of a dark energy component with
negative pressure in addition to other ordinary components as matter and/or
radiation. The most accepted candidate to describe this component is the
scalar field with a negative effective pressure. We can differentiate two
kinds of models, the usual of quintessence and that of k essence with non
canonical kinetic term. These models are different in several aspects, for
instance, in the dynamics of the equation of state and in the behavior of the
sound speed. Precisely, we concentrate our investigations in this last issue. In
quintessence models the scalar field obeys a nearly constant equation of state
with a barotropic index $\gamma_\phi<2/3$, so that $c^2_{s\phi}<0$, but as shown in \cite{Hu} this does not go
against the stability of the
perturbations, because what really matters
is $c_{\rm{eff}}^2$.  In contrast, if the dark energy component is described by
k essence the sound speed (\ref{cs}) could be $c^2_{s}>1$, even
$c^2_{s}=\infty$ is possible as it was seen in previous sections. This means
that perturbations of the background k field can travel faster than light as
measured in the preferred frame where the background field is homogeneous.
This problem should not come as a surprise, because if we calculate the 
adiabatic sound speed we find 

\be
\n{ca}
c_{s\phi}^2=\frac{\dot p_\phi}{\dot\rho_\phi}=-1-\frac{\beta}{\alpha}\dot\phi-
\frac{\ddot\phi}{3H\dot\phi}
\ee
\no we find it is not defined. This happens because in the field equation 
(\ref{kg}), the term $\ddot\phi/c_s^2$ vanishes (due to the fact that the  speed of sound $c_s^2$, as usually  defined in the framework of k-essence, is 
divergent as can be  seen from (\ref{cs})). In consequence, $\dot\phi$ and $\ddot\phi$ are not controlled by any field equation. In other words, the divergent theory
enforces the inverse square potential but the behavior field will come from
additional {\it ad hoc} assumptions.  This makes it evident that for this particular 
particular theory, which is generated by the function (\ref{divergent}), the value of the speed depends strongly on the definition we use to calculate it.
In fact, if we had used the definition given by  Eq. (\ref{ca}) then the speed of sound would, in principle, not have been  divergent an the  field equation
(\ref{kg}) would have been the same as before, and the results obtained in the previous sections would have remained valid.  However, things change radically when a matter 
component is introduced, like in the example discussed before with the help of equations 
(\ref{00m})-(\ref{cm}). In this case the potential is not fixed by the field equation (\ref{kg}) because $\gamma_\phi=-2\dot H/3H^2$ is no longer valid. In this case, the potential has to be assumed independently, and the dynamical equations will fix the field and its derivatives so that the adiabatic sound speed (\ref{ca}) will be perfectly defined. In this case, then, it will be the effective $c_{\rm{eff}}^2$ what we  will have to calculate. To that end we follow the steps of
\cite{Hu}-\cite{perrota}.

In generalized dark matter \cite{Hu} it was introduced the effective sound
speed $c^2_{\rm{eff}}$ defined in the rest frame of the generalized dark matter
component, where $\delta T^0_{j\phi}=0$. The effective sound speed can be
interpreted as a rest frame sound speed, allowing us to define a stabilization
scale for a perturbation, given by the corresponding effective sound horizon.
So, it was possible to show that density perturbation in ordinary quintessence
scenario are damped out below the horizon and the effective sound speed of
quintessence recovers its relativistic behavior $c^2_{\rm{eff}}=1$ \cite{Hu}. In
Ref. \cite{perrota} it was shown that in extended quintessence scenarios
things can be different, because the effective sound speed may be strongly
affected by the nominimal coupled scalar field.

The effect of the speed of sound on the CMB perturbation equations is such
that for an effective sound speed $c_{\rm{eff}}^2\ll 1$ \cite{muk}, k-essence
energy density perturbations are enhanced by perturbations in the cold dark
matter. The perturbed FRW line element is

\begin{equation}
ds^2 = a^2(\eta)[ d\eta^2 - (\delta_{i j} + h_{i j})dx^i dx^j],
\end{equation}

\no where $\delta_{ij}$ is the background spatial metric, $h_{ij}\ll 1$ is the
metric perturbation (we consider only linear metric cosmological
perturbations), $h$ represents the trace of the spatial metric
perturbation, and the synchronous gauge is being used. We investigate the effects due to adiabatic perturbations to the
$k$-essence stress-energy in the synchronous gauge for a mode with wave number
$k$. Besides, we omit the argument $k$ in the amplitude of the perturbation
quantities in the Fourier space. For a generic component in the model
investigated above, it is convenient to separate out the non-adiabatic entropy
contributions. Hence, for the $k$ field we have

\be
\n{1}
p_\phi\Gamma_\phi=\delta p_\phi -c_{s\phi}^2\delta\rho_\phi,
\ee

\no where $c_{s\phi}^2$ is the adiabatic sound speed (\ref{ca}). Following
Ref. \cite{Hu}, we can write the gauge invariant entropy term as

\be
\n{2}
(1+\gamma_\phi)\Gamma_\phi=\left(c_{\rm{eff}}^2-c_{s\phi}^2\right)
\frac{\delta^{(rest)}\rho_\phi}{\rho_\phi}.
\ee

\no The gauge transformation into an arbitrary frame gives the density
contrast in the dark energy rest frame \cite{Hu}
                        
\be
\n{3}
\frac{\delta^{(rest)}\rho_\phi}{\rho_\phi}=\frac{\delta\rho_\phi}{\rho_\phi}
+3{\cal H}\gamma_\phi\frac{v_\phi-B}{k},
\ee

\no yielding a manifestly gauge-invariant form for the nonadiabatic entropy
contribution \cite{bar}-\cite{ko}. In this equation ${\cal H} \equiv
\dot{a}/{a}$, where overdots represents differentiation with respect to the
conformal time $\eta=\int dt/a$.

Combining Eqs. (\ref{1})-(\ref{3}) with the equation of state for the k field
$p_\phi=-(1+\beta\dot\phi/\alpha)\rho$, we obtain

\begin{equation}
c_{\rm{eff}}^2=\frac{\delta p_\phi-
3\beta{\cal H}V\dot\phi c_{s\phi}^2(v_\phi-B)/k}
{\delta\rho_\phi-3\beta{\cal H}V\dot\phi c_{s\phi}^2(v_\phi-B)/k}.
\label{csk}
\end{equation}

\no From this equation, we conclude that $c^2_{\rm{eff}}\approx \delta
p_\phi/\delta\rho_\phi$ on scales approaching the horizon. The overall effect
is that the pressure fluctuations $\delta p_\phi$ are weak and $k$-essence
perturbations are enhanced via gravitational instability of the matter field.
Finally, for our divergent sound speed model we obtain that on subhorizon
scales

\be
\n{fin}
c^2_{\rm{eff}}\approx\frac{\delta p_\phi}{\delta\rho_\phi}
\approx\gamma_\phi-1+\frac{V}{V'}\frac{d\gamma_\phi}{d\phi},
\ee

\no where $\gamma_\phi$ is associated with $\gamma_m$ and the geometry
through Eq. (\ref{ov}). We have also used that the dynamical equations (\ref{00m})-(\ref{cm}) fix the derivatives of the field once the potential has been given
so that $\dot\phi=\dot\phi(\phi)$, so that $\delta
x=-2\dot\phi\delta\phi d\dot\phi/d\phi$.

It is interesting to study the approximate expression for the finite sound
speed $c^2_{\rm{eff}}$ given by Eq. (\ref{fin}). To this end we investigate a
simple example in which the barotropic index has the form
\be
\n{gat}
\gamma_\phi=2\ln{\frac{\phi_0}{\phi}},
\ee

\no where $\phi_0$ is a constant, so that for an inverse square potential
\be
\n{ct}
c^2_{\rm{eff}}=\gamma_\phi.
\ee

\no An accelerated dark energy scenario requires that $\gamma_\phi<2/3$,
consequently for this model the value of the sound speed satisfy the condition
$c^2_{\rm{eff}}<1$. Combining Eqs. (\ref{ga1}) with (\ref{gat}) we can investigate
the solutions of the remaining equation near the value of the k field
$\phi=\phi_0$. Writing $\phi=\phi_0+\epsilon$ with $|\epsilon|\ll 1$, we find
that to first order in $\epsilon$

\be
\n{e.}
\dot\epsilon=\frac{2\alpha}{\beta\phi_0}\epsilon.
\ee

\no Thus, whenever $\alpha(\beta\phi_0)^{-1}<0$ the constant solution
$\phi=\phi_0$ is stable, then the barotropic index and the sound speed have a
vanishing limit for large cosmological times.

\section{Conclusions}

This work means to contribute to a better understanding of a divergent speed
of sound in k-essence cosmological models. We have  reviewed  some  known
results by considering a theory of that kind, for which the Lagrangian density
depends linearly on the time derivative of the k-field. In previous works,
solutions were obtained under the hypothesis of an inverse square potential.
In contrast, we have shown here that for the theory to be consistent the
potential of the k-field cannot take any other form. Then, following with
revision  we have reminded the  known result that the corresponding power-law
solutions can be mapped to power-law solutions of theories with no divergence
in the speed of sound.

After that, we have presented two important new results that reinforce our
view that k-essence cosmologies with a divergent speed of sound are very
special indeed. First, we have constructed a detailed argument that shows that
the requirement of a divergent sound speed at some point fixes uniquely the
form of the Lagrangian of the theory to be exactly the one considered earlier.
After that we have shown that in the divergent theory the  Hubble factor
depends on the k-field only, whereas in the non-divergent ones depends on its
first derivative too. On the other hand, we have proved that, from the
cosmological point of view, the most interesting solutions belonging to the
divergent theory are asymptotically stable.

We then have consider matter in addition of the k essence studied in the  previous section.
It turns out that in this alternative model k-essence the potential is not fixed any longer, and may be freely chosen. This also brings implications on the admissible definitions of the sound speed. Following in this respect \cite{Hu} and \cite{perrota} 
we have studied the behavior of linear perturbations. The dark energy
clustering in k essence scenarios, where the k field is assumed to be
responsible for the cosmic acceleration today can be realized. The scalar
field density perturbations can grow on sub-horizon scales and the effective
sound speed $c_{\rm{eff}}^2$ may satisfy the requirement $c_{\rm{eff}}^{2}\ll 1$ for a
large set of potentials. In particular, we have introduced a cosmological model
for which the effective sound speed and barotropic index have both suitable
asymptotic properties.

\section*{Acknowledgments}

L.P.C. is partially funded by the University of Buenos Aires  under
project X223, and the Consejo Nacional de Investigaciones Cient\'{\i}ficas y
T\'ecnicas.  R.L. is supported by  the University of the Basque Country through research grant 
UPV00172.310-14456/2002, by the Spanish Ministry of Science and Technology through research grant  BFM2001-0988, and  by the Basque Government through fellowship BFI01.412. 

\end{document}